\documentclass[twocolumn,aps,prl,superscriptaddress,longbibliography]{revtex4-2}
\usepackage{amssymb,bm}
\usepackage{graphicx}
\usepackage{amsmath}
\usepackage{braket}
\usepackage{bbm}
\usepackage[colorlinks=true,linkcolor=blue,citecolor=blue,urlcolor=blue]{hyperref}
\usepackage{xcolor,soul}

\begin{document}

\title{Trapping $11{,}000$ Atoms in a Tweezer Array Generated by a Single Metasurface}
\author{Yuqing Wang}
\thanks{These authors contributed equally to this work.}
\affiliation{Department of Physics and State Key Laboratory of Low Dimensional Quantum Physics, Tsinghua University, 100084, Beijing, China.}
\author{Zhongchi Zhang}
\thanks{These authors contributed equally to this work.}
\affiliation{Qosmos Co., Ltd., Building 1, Number 7, Jinghui East Road, Beijing Economic and Technological Development Zone, 100176, Beijing, China.}
\affiliation{Department of Physics and State Key Laboratory of Low Dimensional Quantum Physics, Tsinghua University, 100084, Beijing, China.}
\author{Tao Zhang}
\thanks{These authors contributed equally to this work.}
\affiliation{Institute for Advanced Study, Tsinghua University, 100084, Beijing, China.}
\author{Yuxuan Liao}
\thanks{These authors contributed equally to this work.}
\affiliation{Department of Electronic Engineering, Tsinghua University, 100084, Beijing, China.}
\author{Hanteng Wang}
\affiliation{Institute for Advanced Study, Tsinghua University, 100084, Beijing, China.}
\author{Ye Tian}
\affiliation{Qosmos Co., Ltd., Building 1, Number 7, Jinghui East Road, Beijing Economic and Technological Development Zone, 100176, Beijing, China.}
\affiliation{Department of Physics and State Key Laboratory of Low Dimensional Quantum Physics, Tsinghua University, 100084, Beijing, China.}
\author{Binjie Ji}
\affiliation{Qosmos Co., Ltd., Building 1, Number 7, Jinghui East Road, Beijing Economic and Technological Development Zone, 100176, Beijing, China.}
\author{Yujia Wu}
\affiliation{Department of Physics and State Key Laboratory of Low Dimensional Quantum Physics, Tsinghua University, 100084, Beijing, China.}
\author{Luming Ma}
\affiliation{Department of Physics and State Key Laboratory of Low Dimensional Quantum Physics, Tsinghua University, 100084, Beijing, China.}
\author{Chen Qing} 
\affiliation{Department of Electronic Engineering, Tsinghua University, 100084, Beijing, China.}
\author{Chengshu Li}
\affiliation{Institute for Advanced Study, Tsinghua University, 100084, Beijing, China.}
\author{Wei Zhang}
\affiliation{Department of Electronic Engineering, Tsinghua University, 100084, Beijing, China.}
\author{Yidong Huang}
\affiliation{Department of Electronic Engineering, Tsinghua University, 100084, Beijing, China.}
\author{Wenjun Zhang}
\email{zhangwj22@mails.tsinghua.edu.cn}
\affiliation{Department of Physics and State Key Laboratory of Low Dimensional Quantum Physics, Tsinghua University, 100084, Beijing, China.}
\author{Xue Feng}
\email{x-feng@tsinghua.edu.cn}
\affiliation{Department of Electronic Engineering, Tsinghua University, 100084, Beijing, China.}
\author{Wenlan Chen}
\email{cwlaser@mail.tsinghua.edu.cn}
\affiliation{Department of Physics and State Key Laboratory of Low Dimensional Quantum Physics, Tsinghua University, 100084, Beijing, China.}
\author{Hui Zhai}
\email{hzhai@tsinghua.edu.cn}
\affiliation{Institute for Advanced Study, Tsinghua University, 100084, Beijing, China.}

\date{\today}

\begin{abstract}

\textbf{The scalability of physical qubit numbers is a central challenge toward a universal fault-tolerant quantum computer. The inherent scalability of atom array quantum computers stems from the identical nature of atomic qubits, so the available qubit resource is primarily limited by the number of atoms that can be trapped and controlled. Here, we robustly trap $11{,}000$ individual atoms in a tweezer array, thereby enabling the available qubit resource to reach the tens-of-thousands scale for the first time among all quantum computation platforms. This advance is enabled by a single metasurface, approximately $2$~cm in diameter, that generates the entire tweezer array without the need for microscope objectives, thereby maximizing laser-power efficiency. The large aperture ensures a working distance of about $1.5$~cm, allowing the metasurface to be placed outside the vacuum cell and avoiding the technical complications of in-vacuum operation. We further characterize the randomly loaded atom array using the statistical theory of percolation phase transitions. This work takes an important first step toward a quantum computer at the $10{,}000$-qubit scale.}

\end{abstract}

\date{\today}
\maketitle

It is widely expected that a practically useful quantum computer will require at least tens of thousands of qubits~\cite{babbush2025grand, beverland2022assessing}. Atom arrays have rapidly emerged as one of the most promising platforms for universal fault-tolerant quantum computing, owing to their excellent scalability~\cite{manetsch2025tweezer, chiu2025continuous, lin2025ai}, high-fidelity quantum gates~\cite{evered2026high, muniz2025high, peper2406spectroscopy, evered2023high, ma2023high}, and the ability to reconfigure qubits for non-local operations~\cite{bluvstein2026fault, bluvstein2024logical}. Recently, it has been proposed that Shor's algorithm could be executed at cryptographically relevant scales using as few as ten thousand atomic qubits, leveraging their reconfigurable advantage~\cite{cain2026shor}. On this platform, the available qubit resource is primarily limited by the total number of individual atoms that can be trapped and controlled by optical tweezer arrays.

Conventional methods for generating tweezer arrays primarily rely on acousto-optic deflectors (AODs) and spatial light modulators (SLMs). These approaches are fundamentally constrained by several critical factors: finite pixel count or resolution, low laser-power utilization efficiency, limited laser-damage thresholds, and narrow diffraction angles that inherently restrict the addressable field of view (FOV). Furthermore, these platforms inevitably require high numerical aperture (NA) microscope objectives for tight focusing. Such objectives introduce additional bottlenecks, including significant optical power attenuation, inherent damage thresholds of the internal optics, and a restricted FOV that limits large-scale expansion. Together, these cumulative constraints typically limit the realizable array size to at most a few thousand traps per device. To date, the best reported result is $6100$ atoms~\cite{manetsch2025tweezer}, achieved by the Caltech group using two commercial SLMs in combination, whereas combining more SLMs dramatically increases the experimental complexity. Therefore, this record for the maximum number of trapped atoms has remained unchanged for more than two years.

Metasurfaces are planar devices typically comprising a transparent substrate patterned with subwavelength nanostructures (meta-atoms) made from high-refractive-index materials (e.g., silicon, SRN, TiO$_2$). These meta-atoms can be individually engineered to deliver full-range phase coverage with subwavelength resolution, enabling arbitrary wavefront shaping~\cite{kildishev2013planar}. Their nanoscale features allow the integration of a much larger number of control pixels within a compact device, generating arrays of up to millions of optical tweezers. Metasurfaces offer ultralow loss, high transmission, and superior laser-damage thresholds. Crucially, the subwavelength spacing of meta-atoms permits extremely sharp phase gradients and steep light deflection, providing a significantly wider addressable FOV. By producing these sharp phase gradients for direct focusing to micrometer-scale spot sizes, metasurfaces can eliminate the need for an additional microscope objective for tweezer array formation~\cite{khorasaninejad2016metalenses}. These capabilities collectively enable metasurfaces to trap tens of thousands, or even more, individual atoms, overcoming the limitations of conventional methods for generating optical tweezers~\cite{holman2026trapping, wang2025direct}.

\begin{figure*}
\begin{center}
    \includegraphics[width=2\columnwidth]{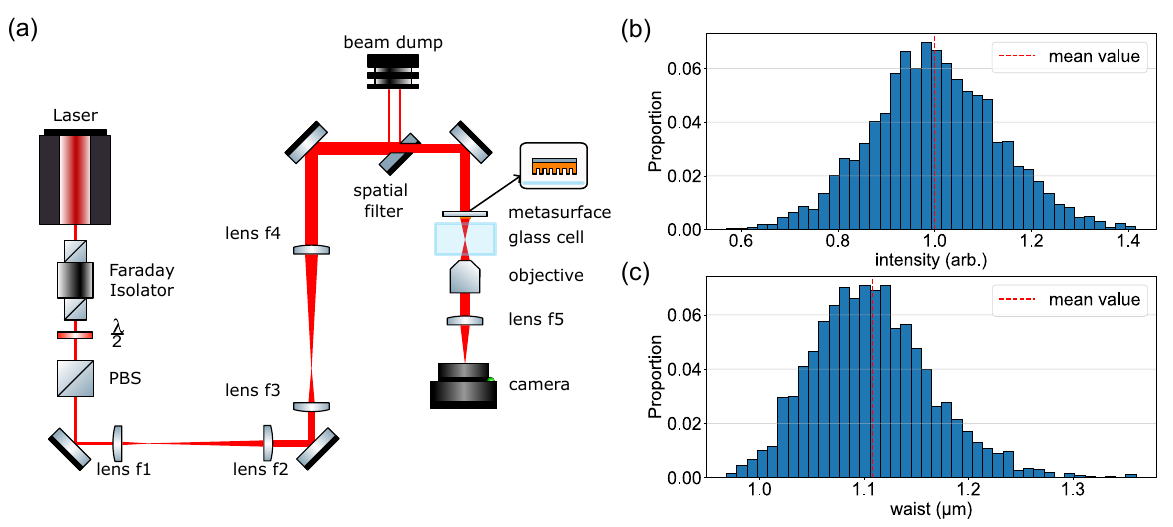}
   \caption{\textbf{Experimental configuration and statistical characterization of the tweezer array.} \textbf{(a)} Schematic of the experimental setup. Key components include a Faraday isolator and polarization optics for beam conditioning, two telescope stages ($f_1$--$f_4$) for beam expansion, a reflective spatial filter to truncate the Gaussian tail, the metasurface as the primary diffractive element, and a monitoring system comprising a high-NA objective, tube lens ($f_5$), and camera. \textbf{(b)} Histogram of the normalized peak intensity distribution of individual tweezers. \textbf{(c)} Histogram of the measured Airy disk radius of tweezers in the focal plane.}
    \label{optical}
\end{center}
\end{figure*}

However, when using a metasurface to generate optical tweezers for atom trapping, the working distance becomes another key consideration. Typically, metasurfaces generate optical tweezer arrays with a working distance of a few millimeters. To trap atoms with such a device, one must place the metasurface inside the vacuum chamber, just a few millimeters away from the trapped atoms~\cite{chen2025multifunctional, li2025vacuum, hsu2022single}. When the metasurface is placed inside the vacuum glass cell, challenges include thermal management issues, electric-field noise that can degrade two-qubit gate fidelities, the technical complexity of ensuring ultra-high vacuum compatibility, and the risk of atom deposition on the nanostructures. The alternative is to place the metasurface outside the vacuum cell and use a pair of high-NA microscope objectives to collimate and focus the tweezer array, thereby extending the working distance but reintroducing the limitations of the microscope system that constrain the tweezer array size~\cite{holman2026trapping, huang2024metasurface}.

In this work, we address this challenge by designing and fabricating a metasurface with a diameter as large as $19.8$~mm. To ensure compatibility with a standard cold-atom experimental setup, the metasurface is positioned $2.5$~mm outside a commercial Pyrex glass cell, with the cell's center as the focal plane. This configuration requires an optical design that accounts for a propagation path comprising $2.5$~mm of air, $3.5$~mm of Pyrex glass, and $9$~mm of vacuum, resulting in a working distance of $1.5$~cm. To trap individual atoms, we target a tweezer radius (Airy disk radius) of $0.9~\mu$m, corresponding to NA of $0.58$. The large working distance of $1.5$~cm and the high NA of $0.58$ together constrain the aperture diameter to $19.8$~mm. This metasurface directly generates a dense array of $18{,}225$ optical tweezers without using a microscope and maximizes laser power efficiency. As a result, it traps about $11{,}000$ individual atoms, enabling the available qubit resource to enter the tens-of-thousands scale.

\begin{figure*}
\begin{center}
    \includegraphics[width=1.8\columnwidth]{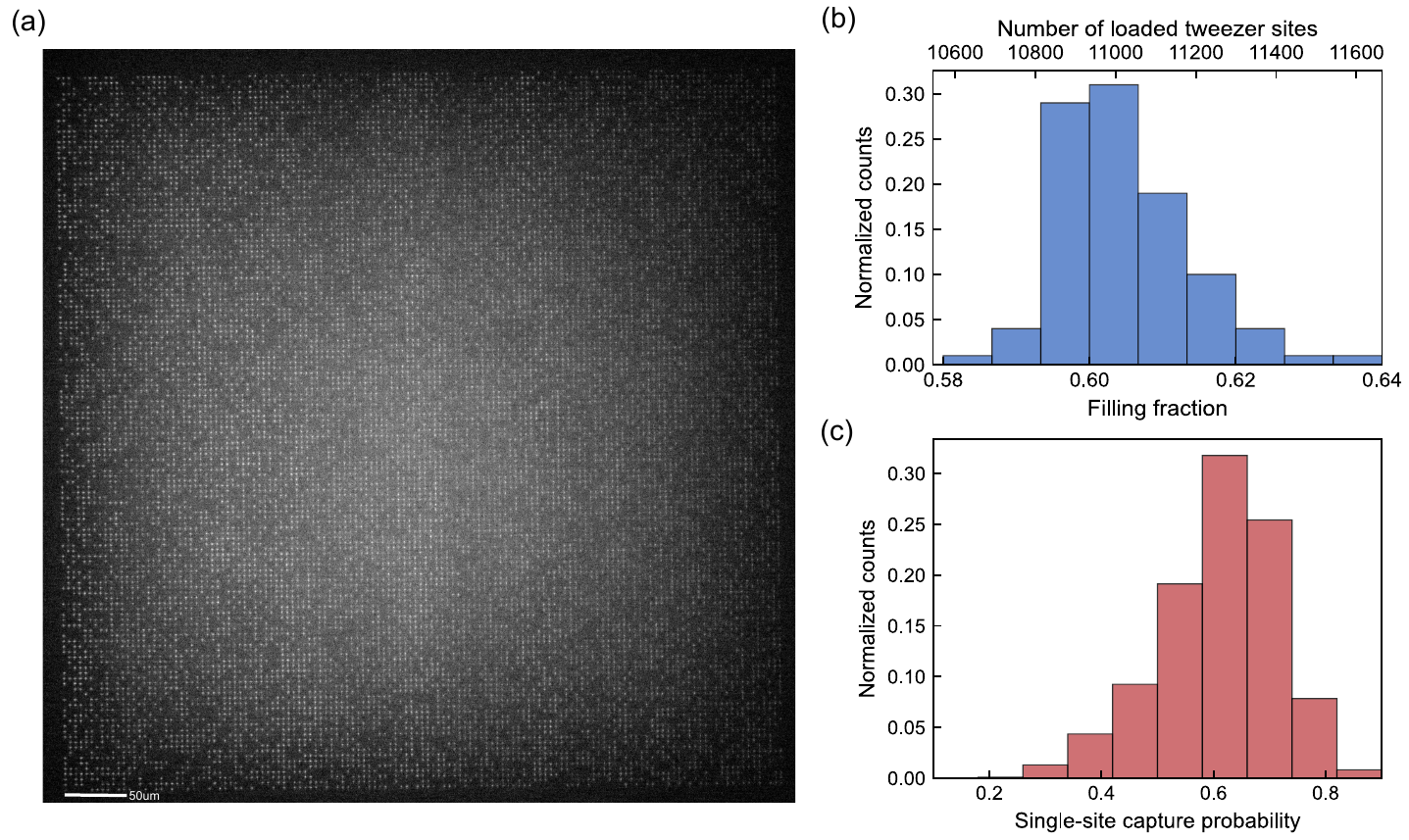}
   \caption{\textbf{Large-scale single-atom trapping performance in the $18{,}225$-site array.} \textbf{(a)} Raw single-shot fluorescence image of $^{87}$Rb atoms stochastically loaded into the $135 \times 135$ metasurface tweezer array. Each bright spot represents a detected single atom; this single shot shows $11400$ trapped atoms. \textbf{(b)} Statistical distribution of the filling fraction over 100 consecutive experimental iterations. The system achieves an average loading of $11{,}022$ atoms with a standard deviation of $\sigma$ = 167, demonstrating high macroscopic stability. \textbf{(c)} Histogram of the site-resolved loading probability across the entire array. The distribution peaks at approximately 60\%.
}
       \label{atoms}
\end{center}
\end{figure*}

\textit{Metasurface design, fabrication, and calibration.} The design and fabrication of a metasurface made from silicon nitride has been described in our previous paper~\cite{wang2025direct}. Now, since the aperture of the metasurface is upgraded to the centimeter-scale, it introduces substantial complexities in both numerical design and nanofabrication. To enable simpler phase calculation, we employ a blocked fast Fourier transform algorithm to efficiently model light field across the large patterned area of the metasurface. The calculation also accounts for refractive-index mismatches among different propagation media to compensate for spherical aberrations and distortions induced by the glass cell. The fabrication processes are optimized to overcome several scale-dependent challenges. Specifically, we mitigate accumulated stitching errors during extended electron-beam lithography due to thermal drift. We also implement a damage-free hybrid stripping process to remove the chromium mask from fragile, high-aspect-ratio silicon nitride nanostructures. Furthermore, we systematically re-optimize reactive ion etching parameters to counteract the macro-loading effect, ensuring morphological consistency and uniform sidewall profiles across the entire $19.8$~mm diameter substrate.

Given the finite laser power available to support the large scale of the tweezer array, we conduct a rigorous analysis of the optical power budget to estimate the achievable trap depth. A critical optimization in this calculation is optimizing the waist of the Gaussian beam incident onto the metasurface, relative to the metasurface aperture. A beam waist significantly smaller than the metasurface radius would underfill the aperture, leading to a reduced effective NA, larger focal spots, and poor array uniformity. Conversely, an excessively large beam waist results in substantial power loss due to truncation. We determine that an incident beam waist approximately equal to the metasurface diameter provides the optimal balance, with only $13.5\%$ of the power lost in the Gaussian tail.
We use a $33$~Watts incident laser, which represents the maximum output power currently available from a single commercial $850$~nm laser (Precilaser FL-SF-850-20-CW). By applying an antireflective coating to the back side of the substrate, reflection losses are effectively eliminated. Consequently, the ideal model predicts a total first-order diffraction power of approximately $17.3$~Watts, accounting for the truncated Gaussian transmission ($86.5\%$), the metasurface diffraction efficiency ($67.5\%$), and the system propagation losses ($10\%$). Besides, because only $87\%$ of the diffracted light forms the effective Airy disk, the effective trapping power is estimated to be $15.0$~Watts. This power budget allocates $0.823$~mW per tweezer, which is designed to provide sufficient trapping potential for robust single-atom loading.

\textit{Experimental setup.} Our experimental setup is shown in Fig.~\ref{optical}(a). The incident 850~nm laser is expanded by two pairs of telescopes to achieve the required collimated beam diameter before illuminating the metasurface. To protect the downstream optomechanics from the Gaussian tail, we intentionally integrate a reflective mirror with a central aperture as a spatial filter, ensuring that only the structured portion of the beam passes. The metasurface itself serves as the only primary focusing element, directly generating the tweezer array in the vacuum glass cell. Notably, the system features a streamlined, single-objective configuration. A high-performance objective is positioned only on the opposite side of the glass cell to monitor the tweezer array and collect atomic fluorescence.

\begin{figure*}
\begin{center}
    \includegraphics[width=1.95\columnwidth]{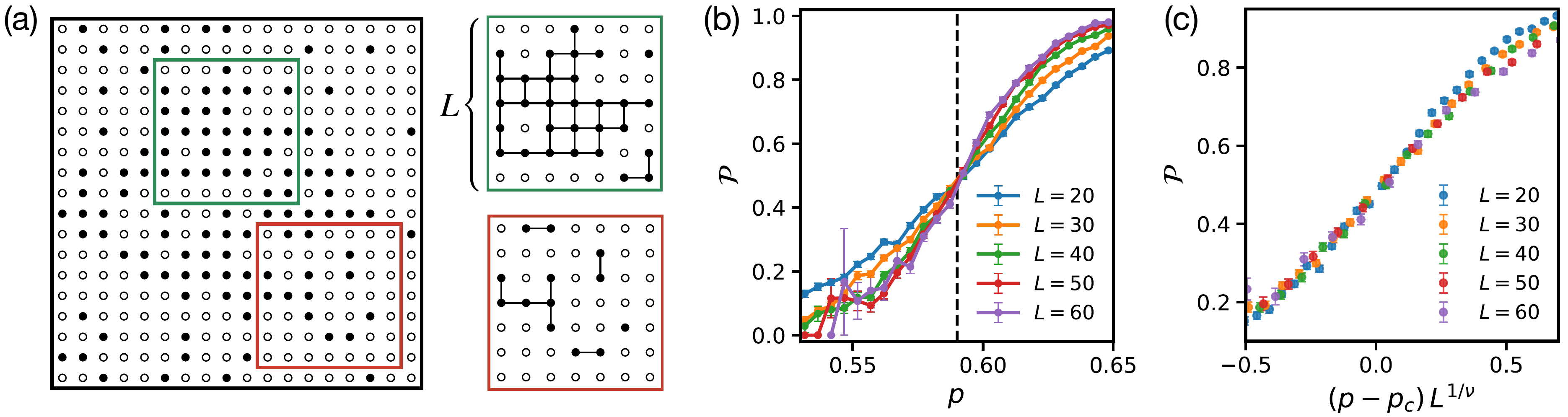}
   \caption{\textbf{Percolation transition from atom-loading snapshots.} \textbf{(a)} Example configurations. Samples of $L \times L$ configurations are chosen within each experimental snapshot. Filled and open circles denote occupied sites and vacant tweezers, respectively, and solid bonds indicate connected links. The green patch illustrates a percolating configuration ($X = 1$); the red patch does not percolate ($X = 0$). \textbf{(b)} Percolation probability $\mathcal{P}(p, L)$ as a function of loading fraction $p$ for different patch sizes $L$. The dashed vertical line marks $p_\text{c} \approx 0.59$. \textbf{(c)} Finite-size scaling collapse plotted against $(p - p_\text{c})L^{1/\nu}$ with $\nu = 4/3$.}    \label{fig:percolation}
\end{center}
\end{figure*}

Using this experimental configuration, we characterize the performance of the generated tweezer array with both low- and high-optical-power measurements. In low-power characterization, the tweezer array exhibits a spacing of $4.30(7)$~$\mu$m and a first dark radius of $1.11(6)$~$\mu$m. This measured radius is slightly larger than the designed $0.9$~$\mu$m, a discrepancy primarily attributed to imperfect beam expansion, residual off-axis aberrations, and fabrication imperfections. The statistical uniformity of the array is further quantified in Fig.~\ref{optical}(b) and (c), which present histograms of the relative intensity distribution of individual tweezers and the beam waist distribution of the Airy disk radius in the focal plane. The measured power non-uniformity across the array is $13.2\%$, demonstrating high consistency across the large-scale array. At full power, the total effective tweezer power is measured as 12.2~Watts, corresponding to a trap depth of 0.302 mK. We attribute the discrepancy between this measured power and the ideally predicted 15.0 Watts mainly to high-power thermooptical effects and alignment sensitivities.

\textit{Results of Trapping Individual Atoms.} We load $^{87}$Rb atoms into the $18{,}225$-site tweezer array from a cold atomic gas. The atoms are sequentially cooled in two-dimensional (2D) and three-dimensional (3D) magnetooptical traps (MOTs). We then load the atoms stochastically into the tweezer array using a D1-line grey molasses stage to ensure single-atom occupation. By overlapping the molasses with the focal plane of the metasurface, we achieve robust parity projection across the entire array.

Following loading, we push out the off-plane atoms and then detect the trapped individual atoms via fluorescence imaging. We illuminate the array with a 780~nm beam for 200 ms and collect the scattered photons using a quantitative CMOS (qCMOS) camera
(Hamamatsu ORCA-Quest 2). Figure 2(a) shows a raw single-shot fluorescence image of atoms randomly loaded in a $135\times135$ tweezer array, demonstrating the simultaneous trapping of $\sim 11{,}000$ single atoms. To quantitatively evaluate performance, we characterize macroscopic stability and array uniformity across 100 consecutive experimental iterations. Globally, our system achieves an average of $11{,}022$ trapped atoms per iteration (Fig. 2(b)), corresponding to a mean filling fraction of 60.5\% across the $18{,}225$ sites. The distribution of the total number of atoms has a standard deviation of only $\sigma$ = 167 atoms, underscoring the stability of our metasurface tweezer array system. To verify the trapping potential, we measure the average trap depth by characterizing the light-induced AC Stark shift. This yields an average trap depth of $0.305(17)$~mK across the array, which matches well with the $0.302$~mK predicted from the 12.2~Watts diffraction power.

At the individual trap level, the site-resolved loading probability histogram (Fig. 2(c)) peaks at approximately 60\%, with a small tail extending towards lower loading rates. This tail is attributed to both systematic variations in trap depth and stochastic atom loss during experimental sequences. These losses are primarily induced by imaging-related heating and the near-resonant pushing procedure used to eliminate off-plane background atoms.

\textit{Percolation Analysis of Randomly Loaded Atom Arrays.} Here we show that the celebrated percolation theory~\cite{StaufferAharony1994} from classical statistical mechanics offers an alternative and more collective diagnosis for randomly loaded atom arrays. A link between two nearest-neighbor sites is defined as \textit{connected} if both tweezer sites are occupied by a single atom. For an $L \times L$ square configuration, the configuration is said to \emph{percolate} if there exists at least one connected path linking two opposite boundaries, either left to right or bottom to top. Different $L \times L$ patches can be chosen within a single experimental snapshot, as illustrated in Fig.~\ref{fig:percolation}(a). The same figure also provides an example: the green patch corresponds to a left-to-right percolating realization, whereas the red patch is non-percolating.

For each configuration $\alpha$, we define a percolation indicator $X_\alpha = 1$ if the configuration percolates, and $X_\alpha = 0$ otherwise. We also assign to the same configuration an average loading fraction $p_\alpha = (1/L^2)\sum_{i \in \alpha} n_i$, where $n_i=1$ or $0$ is the number of atoms in the site, and the sum runs over all sites inside patch $\alpha$. Thus, each configuration is characterized by $(p_\alpha, X_\alpha)$. For a target loading $p$, the percolation probability $\mathcal{P}$ is defined as
\begin{equation}
\mathcal{P}(p, L) = \frac{\sum_{\alpha} X_\alpha \, \delta(p_\alpha - p)}{\sum_{\alpha} \delta(p_\alpha - p)},
\label{eq:percolation_prob}
\end{equation}
where the summation over $\alpha$ includes all patches within each snapshot and across all $100$ experimental snapshots. In practice, $\delta(p_\alpha - p)$ is replaced by a restriction $|p_\alpha - p| < \epsilon$, with $\epsilon = 0.01$ in Fig.~\ref{fig:percolation}(b) and (c).

For a uniformly random loading, percolation theory predicts a second-order phase transition at $p_\text{c} \approx 0.59$~\cite{Ziff1992}. For $p < p_\text{c}$, the system does not percolate, with $\mathcal{P}(p, L=\infty) = 0$, and $\mathcal{P}(p, L=\infty)$ becomes non-zero for $p > p_c$. The theory further predicts that, for finite $L$, all $\mathcal{P}(p, L)$ near the critical point should obey a universal scaling form $f[(p-p_c)L^{1/\nu}]$~\cite{Cardy1996, Stauffer1979}. Consequently, all $\mathcal{P}(p, L)$ curves with different $L$ are expected to cross at $p_\text{c}$, as shown in Fig.~\ref{fig:percolation}(b). For the percolation transition, the critical exponent $\nu = 4/3$~\cite{Isichenko1992}, which is verified in Fig.~\ref{fig:percolation}(c) by the collapse of the data near $p_\text{c}$ when $\mathcal{P}(p, L)$ is plotted against $(p-p_c)L^{3/4}$.

\textit{Outlook.} In summary, we have demonstrated that direct projection of a tweezer array by a single metasurface placed outside the vacuum cell can trap about $11{,}000$ individual atoms. We note that this total number of atoms is by no means the limit of this technology development. By combining the power of two $850$~nm lasers and, in the near future, doubling the laser power, we estimate that the number of trapped atoms could increase by approximately a factor of $3-4$ without reaching the laser damage threshold. The total number of atoms can also be increased by choosing an $830$~nm laser with a smaller detuning of the atomic resonance, which provides a stronger trapping potential at the same laser power. Our approach can be directly applied to other atomic species, such as Cs, where a $1064$~nm laser can reach a power of $300$~Watts, potentially increasing the total number by a factor of approximately $4-6$ even with a single laser.

We further demonstrate that percolation theory provides a robust description of a randomly loaded atom array. Near the critical point, percolation theory predicts intriguing features such as fractal geometry. Even without rearrangement, it is intriguing to study many-body correlations arising from the Rydberg blockade effect in such fractal structures. To rapidly rearrange large-scale arrays of atoms, an artificial-intelligence-based algorithm is essential~\cite{zhang2026algorithm, lin2025ai}. Recent algorithmic advances have enabled the rearrangement of near-perfect arrays of approximately $10{,}000$ atoms within about $20$~ms~\cite{zhang2026algorithm}. The next milestone will be to assemble a defect-free atom array at this scale, ultimately paving the way towards fully controlled qubit arrays at the $\sim 10{,}000$ qubit scale.

\textbf{Acknowledgments.} This work was supported by the National Natural Science Foundation of China (Grant No. 92576208), Tsinghua University Initiative Scientific Research Program, Beijing Science and Technology Planning Project, and Tsinghua University Dushi Program.

\bibliography{Ten-thousand-individual-atoms} 

\end{document}